\documentclass[fleqn,usenatbib]{mnras}

\usepackage{newtxtext,newtxmath}

\usepackage[T1]{fontenc}

\DeclareRobustCommand{\VAN}[3]{#2}
\let\VANthebibliography\thebibliography
\def\thebibliography{\DeclareRobustCommand{\VAN}[3]{##3}\VANthebibliography}

\usepackage{xcolor}
\definecolor{OliveGreen}{RGB}{50, 205, 50}

\DeclareMathOperator*{\argmin}{argmin}

\newcommand{\norm}[1]{\left\lVert#1\right\rVert}
\newcommand{\isep}{\mathrel{{.}\,{.}}\nobreak}

\usepackage{graphicx}	
\usepackage{amsmath}	
\usepackage{url}

\title[ML-GPR for EoR]{Retrieving the 21-cm signal from the Epoch of Reionization with learnt Gaussian process kernels}

\author[F.G. Mertens et al.]{
Florent G. Mertens,$^{1,2}$\thanks{E-mail: florent.mertens@obspm.fr}
Jérôme Bobin,$^{3}$
Isabella P. Carucci$^{4,5}$
\\
$^{1}$LERMA, Observatoire de Paris, PSL Research University, CNRS, Sorbonne Université, F-75014 Paris, France\\
$^{2}$Kapteyn Astronomical Institute, University of Groningen, PO Box 800, 9700AV Groningen, The Netherlands\\
$^{3}$IRFU, CEA, Universite Paris-Saclay, F-91191, Gif-sur-Yvette, France\\
$^{4}$INAF - Osservatorio Astronomico di Trieste, Via G.B. Tiepolo 11, 34131 Trieste, Italy\\
$^{5}$IFPU - Institute for Fundamental Physics of the Universe, Via Beirut 2, 34151 Trieste, Italy
}

\date{Accepted XXX. Received YYY; in original form ZZZ}

\pubyear{2023}

\begin{document}
\label{firstpage}
\pagerange{\pageref{firstpage}--\pageref{lastpage}}
\maketitle

\begin{abstract}
Direct detection of the Cosmic Dawn and Epoch of Reionization via the redshifted 21-cm line of neutral Hydrogen will have unprecedented implications for studying structure formation in the early Universe. This exciting goal is challenged by the difficulty of extracting the faint 21-cm signal buried beneath bright astrophysical foregrounds and contaminated by numerous systematics. Here, we focus on improving the Gaussian Process Regression (GPR) signal separation method originally developed for LOFAR observations. We address a key limitation of the current approach by incorporating covariance prior models learnt from 21-cm signal simulations using Variational Autoencoder (VAE) and Interpolatory Autoencoder (IAE). Extensive tests are conducted to evaluate GPR, VAE-GPR, and IAE-GPR in different scenarios. Our findings reveal that the new method outperforms standard GPR in component separation tasks. Moreover, the improved method demonstrates robustness when applied to signals not represented in the training set. It also presents a certain degree of resilience to data systematics, highlighting its ability to effectively mitigate their impact on the signal recovery process. However, our findings also underscore the importance of accurately characterizing and understanding these systematics to achieve successful detection. Our generative approaches provide good results even with limited training data, offering a valuable advantage when a large training set is not feasible. Comparing the two algorithms, IAE-GPR shows slightly higher fidelity in recovering power spectra compared to VAE-GPR. These advancements highlight the strength of generative approaches and optimise the analysis techniques for future 21-cm signal detection at high redshifts.

\end{abstract}




\section{Introduction}

Observation of the redshifted 21-cm transition line of neutral hydrogen from the Cosmic Dawn (CD; $15 < z < $30) and the Epoch of Reionization (EoR; $z \sim 6 - $15) promises to shed new light on the first billion years of our Universe. It will explore the properties of the first stellar population, the detailed timing of the heating and reionization era, as well as the sources that drove them~\citep[see e.g.,][for extensive reviews]{Pritchard12,Furlanetto16}. 

Many observational programs are underway to detect the 21-cm signal. Current radio interferometers (LOFAR\footnote{Low-Frequency Array, http://www.lofar.org}, MWA\footnote{Murchison Widefield Array, http://www.mwatelescope.org}, NenuFAR\footnote{New Extension in Nancay Upgrading LOFAR, https://nenufar.obs-nancay.fr/en/homepage-en/}, HERA\footnote{Hydrogen Epoch of Reionization Array, http://reionization.org}) will provide a statistical characterization of this signal and are paving the way to the next generation SKAO\footnote{SKA Observatory, https://www.skao.int/} that promises an exceptional gain in sensitivity, allowing one to obtain images of these distant epochs, and in particular of the ionized hydrogen bubbles, sources of incomparable information on the physical conditions governing the EoR~\citep{Koopmans15}.

Detecting the 21-cm signal is extremely challenging, mainly due to the difficulty of extracting this very faint signal buried beneath astrophysical foregrounds several orders of magnitude brighter and contaminated by numerous systematics. Although the emission mechanisms of the foregrounds are well known, primarily synchrotron and free-free emission, the intrinsic chromatic response of the instrument and calibration errors create an additional impediment, introducing rapid spectral variations to otherwise spectrally regular foregrounds~\citep[e.g.,][]{Thyagarajan15,Ewall17}. Mitigating these sources of chromaticity is complex. Many experiments have adopted a so-called `foreground avoidance' strategy, which only looks for the signal inside a region in $k$-space where the thermal noise and 21-cm signals dominate~\citep[e.g.][]{Jacobs16,Kolopanis19}, discarding the so-called foregrounds-wedge part where the foregrounds dominate, but at the cost of a reduced sensitivity (a factor $\sim$ 3 for HERA, and even a factor $\sim$ 6 for LOFAR). In practice, also, leakage above the wedge is often observed~\citep[e.g.][]{Patil16,Offringa19a,kern19a}. 

To take full advantage of the sensitivity of the instruments, foreground removal algorithms have been developed, particularly algorithms that do not depend on a precise foreground model, which is often difficult to obtain. For instance, Blind Source Separation (BSS) methods have been applied in the EoR context~\citep[e.g.,][]{Chapman12,Chapman13,Bonaldi15,Patil17}. BSS methods linearly decompose data in a set of components in pixel/visibility space modulated in frequency, only assuming statistical properties of the components (e.g., independence or sparsity). For a recent review and comparison of the most used BSS methods, see \cite{Spinelli21}. 

\cite{Mertens18} introduced a method based on Gaussian Process Regression (GPR) that has proven successful in various 21-cm experiments~\citep{Mertens20,Gehlot19,Gehlot20,Ghosh20}. This approach involves constructing a statistical model prior encompassing all components contributing to the observed signal, including foregrounds, mode mixing components, 21-cm cosmological signal, and noise. The method exploits the distinct spectral signatures of these components by utilising parameterised frequency-frequency covariance functions. Initially, the GPR method employed generic analytical covariance models commonly used in Gaussian Processes literature. However, with the increasing integration time of LOFAR observations and the prospect of an SKAO CD/EoR survey, the need for improved accuracy has driven the development of adapted and more realistic covariance models. Moreover, an excessive mismatch of the prior covariance to the true covariance may also bias the recovered 21-cm signal~\citep{Kern21}. Here lies the motivation and objective of the present work.

Analytical formulations of foreground covariance have been developed~\citep{Trott16b,Murray17}. However, they are incomplete, not including, for example, the anisotropy of the galactic emission or the full complexity of the instrument's primary beam. Concerning the 21-cm signal, analytical simulations are far too elementary, and numerical or semi-analytical codes simulating this signal are often incompatible with each other in their parametrisations. Simulations would be more suitable to provide a more physically motivated covariance, whether for the foregrounds, the systematics or the 21-cm signal. The task at hand is thus to transform a set of simulations into a covariance function, able to reproduce the covariance matrices of these simulations and replace the generic analytical covariance models used until now. Here, we consider two machine learning (ML) methods for doing so, Variational Autoencoders~\citep[VAE,][]{Kingma13} and Interpolatery Autoencoders~\citep[IAE,][]{BobinIAE2023}. We train the two with the same set of simulations and compare their performances. Finally, we test both against the \cite{Mertens18} standard GPR pipeline. 

The paper is structured as follows. Section~\ref{sec:methods} reviews the Gaussian Process Regression framework and introduces a generalisation to learnt kernels based on VAE and IAE. Section~\ref{sec:simulations} describes the 21-cm and foregrounds simulated data sets used to test the new algorithm. Section~\ref{sec:results} reports on the results of these simulations, emphasizing the comparisons between autoencoder-based and standard GPR in various scenarios with increasing complexity.

\section{Methods}
\label{sec:methods}

After reviewing the GPR framework, this section describes two different learning-based algorithms to build a covariance model of the 21-cm cosmological signal using simulations: VAE and IAE, Variational and Interpolatory Autoencoders. The combination of the GPR algorithm to perform signal separation with ML-trained kernel is later called ML-GPR.


\subsection{Gaussian Process Regression}
\label{sec:gpr}

In the context of 21-cm signal detection, we are interested in modelling our data $\mathbf{d}$ observed at frequencies $\mathbf{x}$ by a foreground, a 21-cm and a noise component $\mathbf{n}$~\citep{Mertens18}:
\begin{equation}
\mathbf{d} = f_{\mathrm{fg}}(\mathbf{x}) + f_{\mathrm{21}}(\mathbf{x}) +
\mathbf{n}.
\end{equation}
The foregrounds and 21-cm signal components are expected to have distinct frequency-frequency covariance; we can then build a covariance prior model of the data composed of a foreground covariance function (also called kernel) $K_{\mathrm{fg}}$, a 21-cm signal covariance function $K_{\mathrm{21}}$, and a noise covariance $K_{\mathrm{n}} = \mathrm{diag}(\sigma_{ \mathrm{n}}^2(\mathbf{x}))$. The joint probability density distribution of the observations $\mathbf{d}$ and the function values $ \mathbf{f}_{\mathrm{21}}$ of the 21-cm signal component at the same frequencies $\mathbf{x}$ are then given by~\citep{Rasmussen05},
\begin{equation}
\left[ \begin{array}{c} \mathbf{d} \\ \mathbf{f_{\mathrm{21}}} \end{array}\right] \sim  
\mathcal{N}\left( \left[\begin{array}{c} 0 \\ 0 \end{array}\right], \left[ 
\begin{array}{cc} K_{\mathrm{fg}} + K_{\mathrm{21}} + K_{\mathrm{n}} & K_{\mathrm{21}} \\ K_{\mathrm{21}} & K_{\mathrm{21}} \end{array} \right] \right)
\end{equation}
using the shorthand $K \equiv K(\mathbf{x}, \mathbf{x})$. The recovered 21-cm signal is then a Gaussian Process, conditional on the data:
\begin{equation}
\label{eq:gpr_fg_fit_distribution}
\mathbf{f}_{\mathrm{21}} \sim \mathcal{N}\left({\cal E}(\mathbf{f}_{\mathrm{21}}), \mathrm{cov}(\mathbf{f}_{\mathrm{21}})\right)
\end{equation}
with expectation value and covariance defined by:
\begin{align}
\label{eq:gpr_predictive_mean_eor} 
{\cal E}(\mathbf{f}_{\mathrm{21}}) &= K_{\mathrm{21}}\left[K_{\mathrm{fg}} + K_{\mathrm{21}} + K_{\mathrm{n}}\right]^{-1} 
\mathbf{d}\\
\label{eq:gpr_predictive_cov_eor}
\mathrm{cov}(\mathbf{f}_{\mathrm{21}}) & = K_{\mathrm{21}} - K_{
\mathrm{21}}\left[K_{\mathrm{fg}} + K_{\mathrm{21}} + K_{\mathrm{n}}\right]^{-1}K_{\mathrm{21}}.
\end{align}
One may also decide to recover the fitted foregrounds component instead, $\mathbf{f}_{\mathrm{fg}}(\mathbf{x})$, and subtract it from the input data to obtain the residual, as is done in~\cite{Mertens18} and~\cite{Mertens20}.

In GPR, we perform prior covariance model selection under a Bayesian framework by choosing the model that maximizes the marginal-likelihood, i.e., the integral of the likelihood times the prior. The functional form of the covariance is first selected, guided by our knowledge of the fluctuations we aim to model. We then use standard optimization or MCMC methods to determine the optimal hyperparameters of the chosen covariance functions, like the coherence-scale or the variance. The marginal-likelihood is given by:
\begin{equation}
p(\mathbf{d}|\mathbf{x}, \theta) = \int{p(\mathbf{d}|\mathbf{f} ,\mathbf{x}, \theta)p(
\mathbf{f}|\mathbf{x}, \theta)d\mathbf{f}},
\end{equation}
with $\theta$ being the hyperparameters of the covariance function. Under the assumption of Gaussianity, we can integrate over $\mathbf{f}$ analytically, yielding the log-marginal-likelihood~(LML),
\begin{equation}
\label{eq:hyper_lml}
\mathrm{log}\,p(\mathbf{d} |\mathbf{x}, \theta) = -\frac{1}{2}
\mathbf{d}^\intercal (K + \sigma_n^2 I)^{-1} \mathbf{d}
- \frac{1}{2} \mathrm{log}\,|K + \sigma_n^2 I| - \frac{n}{2} \mathrm{log}\,2\pi
\end{equation}
with $n$ the number of sampled points. The posterior probability density of the hyperparameters is then found by applying Bayes' theorem, incorporating here the prior on the hyperparameters:
\begin{equation}
\label{eq:hyper_post}
\mathrm{log}\,p(\theta| \mathbf{d}, \mathbf{x}) \propto \mathrm{log}\,p(\mathbf{d} |\mathbf{x}, \theta) + 
\mathrm{log}\,p (\theta).
\end{equation}
The Matérn class of covariance functions are commonly used as prior covariance for the different components of the data. It is defined by,
\begin{equation}
\label{eq:matern_cov}
\kappa_{\mathrm{Matern}}(r) = \frac{2^{1 - \eta}}{\Gamma(\eta)}\left(
\frac{
\sqrt{2\eta}r}{l}\right)^{\eta}
K_{\eta}\left(\frac{\sqrt{2\eta}r}{l}\right),
\end{equation}
where $K_{\eta}$ is the modified Bessel function of the second kind. Functions obtained with this kernel class are at least $\eta$-time differentiable.  Spectrally-smooth components (typically foregrounds) are best modelled with higher $\eta$, while for the 21-cm signal, we set $\eta = 1/2$ or $\eta = 3/2$~\citep{Mertens18}. Component separation is most effective when the coherence-scale hyperparameter $l$ of the foregrounds covariance is significantly larger than that of the 21-cm signal covariance. However, generic covariance functions may not always accurately represent the true covariance and in the low signal-to-noise regime, an inaccurate prior covariance can introduce biases in the results~\citep{Kern21}. To address this, we can enhance the prior covariance model by learning from simulations, allowing for a more accurate representation.

\subsection{Learning covariance prior model from simulations}
\label{sec:learning_cov}

We assume we have access to $T$ simulations of the 21-cm cosmological signal that allows building a training set of covariances: $\mathcal{T}_\mathrm{train} =  \left \{K^t \right \}_{t = 1,\cdots,T}$. The prime goal of learning is to derive a low-dimensional representation or model of these covariances. From the mathematical modelling point of view, physically relevant covariances generally evolve smoothly, meaning they are likely to belong to a {\it low-dimensional manifold}. When derived from parametric models, the dimensionality of this manifold is exactly the number of free parameters in the model. However, we notice that the dependency between the underlying parameters and the measured covariances is generally {\it non-linear} and either unknown or challenging to handle in a component separation process. Consequently, one has to resort to learning a non-linear model of the covariances to build an efficient and reliable model, relying on an unsupervised manner.

\subsubsection{Variational Autoencoders}
\label{sec:vae}

Among machine learning strategies, autoencoders~\citep[AE,][]{Hinton06} represent the most straightforward dimensionality reduction technique. 
In our context, the Variational Autoencoder ~\citep[VAE,][]{Kingma13} is particularly well adapted since it allows building a potentially complex statistical description of the data.

More formally, the set of covariances $K$ is statistically characterised by some unknown probability distribution $p_\psi(K)$, which depends on some parameters $\psi$. This distribution can be described as the marginalised distribution of the joint distribution $p_\psi(K,z)$ with respect to some latent parameters $z$. In practice, the latent parameters will represent a low-dimensional description of the covariances. The marginalized distribution is obtained as follows:
\begin{equation}
    p_\psi(K) = \int_z p_\psi(K|z) p_\psi(z) dz
\end{equation}
where $p_\psi(z)$ is the prior distribution of the latent variable and $p_\psi(K|z)$ is the conditional distribution of the covariances with respect to $z$. This equation defines how the distribution of $K$ can be derived from the latent variables, eventually defining a generative model for the covariances.\\
In practice, the training stage corresponds to learning the parameters $\psi$ that define the above distributions. To do so, we need a description of the conditional distribution of the latent variables with respect to $K$, which requires the knowledge of $p_\theta(z|K)$. However, this quantity is generally unknown and cumbersome to compute. In the framework of the VAE, it is approximated by a surrogate distribution $q_\phi(z|K)$, where $\phi$ stands for some parameters to be learnt.\\
The training stage then boils down to learning the parameters $\psi$ and $\phi$ so that   $q_\phi(z|K)$ is as close as possible to $p_\theta(z|K)$ according to the Kullback-Leibler divergence $\mathcal{D}_{KL}$. The learning loss eventually writes as~\citep{Kingma:2014aa}:
\begin{equation}
    \min_{\phi,\psi} -\mathbb{E}_{q_\phi}\{\log(p_\psi(K|z))\} + \mathcal{D}_{KL} \left( q_{\phi}(z|K), p_{\psi}(z) \right) 
\end{equation}

Essentially, the VAE can be described with the following ingredients:
\begin{itemize}
    \item{\bf The encoder:} The distribution $q_{\phi}(z|K)$ defines a statistical encoder that samples the latent space conditionally to the covariances. The encoder builds upon a neural network $\Phi$ to describe $q_{\phi}(z|K)$, and the parameters $\phi$ stand for the network parameters. The latent variable is then defined as follows:
    $$
    \forall t = 1,\cdots,T; \quad z_t = \Phi(K^t)
    $$
    
    \item{\bf The decoder:} As stated earlier, the distribution of the covariances $p_\psi(K)$ can be defined as the marginalization with respect to $z$ of the joint distribution $p_\psi(K,z) = p_\psi(K|z) p_\psi(z)$. The prior distribution is generally chosen as a normal distribution, whose mean and covariance matrix are trained. Sampling the distribution $p_\psi(K)$ is then obtained by first drawing from the prior distribution of the latent variable: $\tilde{z}^t \sim p_\psi(z)$ and then applying a non-linear function $\Psi$ to map it back to the input domain:
    $$
    \forall t = 1,\cdots,T; \quad \hat{K^t} = \psi(\tilde{z}^t)
    $$
    the so-called decoder $\Phi$ is generally described with a neural network, whose parameters $\phi$ are learnt during training.
\end{itemize}

Details of the VAE implementation we used in the experiments are in Appendix~\ref{app:details}.

\subsubsection{Interpolatory Autoencoder}
\label{sec:iae}

As mentioned above, autoencoders are powerful architectures to build low-dimensional non-linear signal approximations. However, their training can be difficult, especially when the training samples are scarce. A way forward to face these limitations is to impose additional constraints and better {\it organise} the latent space. In this context, the interpolatory autoencoder~\citep[IAE,][]{BobinIAE2023}\footnote{\href{https://github.com/jbobin/IAE}{https://github.com/jbobin/IAE}} aims to capture the geometrical structures of the space in which the data live, which is generally well approximated by some (unknown) low-dimensional manifold $\mathcal{M}$.\\
In contrast to standard AE architectures, IAE maps its latent space with so-called anchor points, which are fixed signal examples known to be on (or close to) the manifold $\mathcal{M}$. These anchor points are generally chosen within the training set. They serve as fixed points from which new samples can be built through a non-linear interpolation scheme.\\
While standard AE looks for a global description of the manifold $\mathcal{M}$, the IAE learns to travel on the manifold $\mathcal{M}$ by interpolating the anchor points.\\
Like the VAE, the IAE architecture is defined with an encoder/decoder scheme. However, a key difference is that IAE decodes low-dimensional linear approximations of the input data in the latent space. The IAE is therefore trained so that the code of each sample can be expressed as a linear combination, or {\it barycenter}, of anchor points in the latent domain. \cite{BobinIAE2023} emphasises how this procedure allows more robustness of the model with respect to the scarcity of the training samples.\\

 We first draw at random $d$ samples from the training $\mathcal{T}$ that will serve as anchor points $\{K_a^{n}\}_{n\in[1\isep d]}$. Similarly to the VAE, we define $\Phi$ and $\Psi$ as the encoder and the decoder, respectively. The IAE networks is then defined by the three following steps:
\begin{enumerate}
    \item{\bf Mapping the data to latent space:} The encoder $\Phi$ maps the data to the latent space, defining some latent vector $\ell^{i} = \Phi(K^{i})$. 
    \item{\bf Interpolation step:} Each sample is approximated as a linear interpolation from the anchor points, i.e., we compute the interpolating  weights $\lambda^{i}$ as the projection coefficients of $\ell^{i}$ onto the vector space generated by the latent codes of the anchor points $\{\Phi(K_a^{n})\}_{n\in[1\isep d]}$:
    \begin{equation}
        \{\lambda[n]\}_{n=1,\cdots,d} = \mbox{argmin}_\lambda \left\| \Phi(K^t) - \sum_{n=1}^d \lambda[n] \Phi(K^{n}_a) \right\|_2^2
    \end{equation}
    The interpolated samples are then defined as:
    \begin{equation}
        \tilde{\ell}^{t} = \sum_{n=1}^d \lambda^{t}[n]\Phi(K^{n}_a) \,.
    \end{equation}
    \item{\bf Mapping the interpolated samples back to the data domain:} The decoder $\psi$ maps back the interpolated samples from the latent space to the data domain: $\tilde{K}^{t} = \Psi(\tilde{\ell}^{t} )$.
\end{enumerate}{}
In this model, both the encoder and the decoder are learnt. Model optimisation is done by minimising the reconstruction error between the training data $\{K^{t}\}_j$ and their approximation, \textit{i.e.}: 
\begin{equation}
	\argmin_{\phi, \psi} \sum_j \norm{K^{t} - \Psi\!\left(\sum_{n=1}^N \lambda^j[n] \Phi\!\left(K_a^{n}\right)\right)}^2_2 \,.
\end{equation}
\\
Once the IAE is learnt, any point of $\mathcal{M}$ can be approximated as an interpolant of the anchor points in the code domain back-projected to the sample domain:
\begin{equation}
	\mathcal{M} \approx \left\{x, \exists \{\lambda[n]\}_{n\in[1 \isep N]} \in \mathbb{R}^N, x = \psi\!\left(\sum_{n=1}^N \lambda[n] \Phi\!\left(K_a^{n}\right)\right)\right\}.
\end{equation}
Details of the architecture we used in the experiments as well as its implementation, are described in Appendix~\ref{app:details}.

\subsubsection{Application of AE-based models for covariance estimation}
\label{sec:learnt_kernel_practice}

AE-based models can be interpreted and applied in different ways. To that respect, it is customary to use AE-based models to compute non-linear, low-dimensional approximations of the classes of signals as sampled by the training set. For instance, once learnt, one can use the IAE model to produce a low-dimensional approximate of some signal $K$ by composing the application of the encoder, the interpolator and the decoder:
$$
\hat{K} = (\Psi \circ I \circ \Phi ) (K) \,.
$$
We can interpret this operation as a projection onto the low-dimensional manifold implicitly defined by the encoder and the decoder.
\\
From a different viewpoint, the decoder of AE-based models boils down to a learnt parametric model for the covariances. Combined with a sampling process, as in the VAE, it leads to generative models.

Let us now consider the case of observations involving the $21$-cm signal along with foreground and noise components. The observation model is described as follows:
$$
K_\mathrm{d} = K_{\mathrm{fg}} + K_{\mathrm{21}} + K_{\mathrm{n}}\,.
$$
In the scope of covariance estimation from observations, one can substitute standard analytic kernels, such as Matérn of Gaussian kernels, with the decoder, which can be treated as any parametric model $\Psi(\lambda)$, where $\lambda$ lives in the latent space of a given AE-based model. The observation model then reads as
$$
K_\mathrm{d} = K_{\mathrm{fg}} +\Psi(\lambda) + K_{\mathrm{n}}\,.
$$
Covariance estimation then boils down to estimating the hyperparameter $\lambda$ during the inference procedure. The {\it a posteriori} distribution now reads as
\begin{equation}
\label{eq:hyper_post2}
\mathrm{log}\,p(\theta, \lambda| \mathbf{d}, \mathbf{x}) \propto \mathrm{log}\,p(\mathbf{d} |\mathbf{x}, \theta,\lambda) + 
\mathrm{log}\,p (\theta)+ 
\mathrm{log}\,p (\lambda),
\end{equation}
where the LML is defined in \autoref{eq:hyper_lml}, and where the hyperparameters $\lambda$ are related explicitly to the parameterisation of the AE-based models that act as the covariance of the $21$-cm signal. The hyperparameters $\theta$ are related to modelling additional components such as noise and foregrounds, depending on the model we consider (see Section~\ref{sec:gpr}).
\\
To compute the maximum a-posteriori (MAP) and estimate uncertainties on the hyperparameters $\theta$ and $\lambda$, we use a Monte Carlo Markov Chain (MCMC). Such methods aim to sample the model parameters' posterior probability distribution given the observed data. In practice, this is performed with the ensemble sampler algorithm of \cite{Goodman10}, implemented in the \texttt{emcee} python package~\citep{Foreman-Mackey13}.

\subsubsection{Training the 21-cm signal kernel}


In the generic description of the learnt kernel, the kernel is trained on a set of covariance $K$. However, when it comes to the specific case of the 21-cm signal, we can exploit the isotropic nature of the signal and train the AE on a set of 21-cm power spectra, denoted as $P_{\mathrm{HI}}(\mathbf{k})$. This training is equivalent to training on the covariance $K$. With the AE, we learn the shape of the power spectra while keeping the variance as a separate parameter. Therefore, during the pre-processing of our learning step, our 21-cm power spectra of the training set are scaled to have the same power. The trained decoder of the AE thus takes $n$ parameters as input; it then outputs a normalized spherically-averaged power spectrum, which is then converted to frequency-frequency covariance matrix, using the expression of the angular power-spectrum, in the flat-sky approximation~\citep{Datta07}:
\begin{equation}
    C_l(\Delta\nu, k_{\perp}) = \frac{T^2}{\pi r^2_{\nu}} \int_{0}^{\infty} P_{\mathrm{HI}}\left(\sqrt{k_{\perp}^2 + k_{\parallel}^2}\right) \cos(k_\parallel r_{\nu} \Delta\nu) \mathrm{d}k_\parallel \,,
\end{equation}
where $T$ is the differential brightness temperature of the 21-cm signal, $r^2_{\nu}$ is the comoving distance at the mean frequency of the observation cube. Additionally, $k_{\perp}$ and $k_{\parallel}$ represent the Fourier modes in the transverse (spatial) and line-of-sight (frequency) directions, respectively. A more comprehensive description of these concepts and their definitions can be found in \autoref{sec:power_spectra_definition}. Combining these elements, we form the 21-cm trained kernel, which takes $n$+1 parameters ($n$ corresponding to the dimension of the latent space of the AE), and one parameter for the variance of the kernel. We note here the dependence of $C_l$ on $k_{\perp}$ which provides a significant advantage over the generic kernel traditionally used in the standard GPR method as later tests will demonstrate.

\subsection{The ML-GPR method}

We refer to the combined approach of the GPR method with a learnt kernel as the ML-GPR method. Similar to the standard GPR method, ML-GPR operates on a gridded visibility cube $V(\mathbfit{u}, \nu)$, which is the Fourier transform along the angular direction of the observed image cube, and where $\mathbfit{u} = (u, v)$ is the vector representing the coordinates in wavelength in the visibility plane and $\nu$ is the observing frequency. This choice aligns with the nature of radio interferometric observations, where visibilities are directly measured. 

\subsection{Power spectra estimation}
\label{sec:power_spectra_definition}

Current generation of redshifted 21-cm experiments aims at characterising the power spectra of the signal at different redshift to quantify the scale dependent second moment of the signal and its evolution as function of redshift. We compute it by taking the Fourier transform of the gridded visibility cubes $V(\mathbfit{u}, \nu)$ in the frequency direction. We can then define the cylindrically averaged power spectrum as~\citep{Mertens18}:
\begin{equation}
P(k_{\perp}, k_{\parallel}) = \frac{X^2 Y}{\Omega_{\mathrm{PB}} B} \left<\left|
\hat{V}
(\mathbfit{u}, \tau)\right|^2\right>,
\end{equation}
where $\hat{V}(\mathbfit{u}, \tau)$ is the Fourier transform in the frequency direction, $B$ is the frequency bandwidth, $\Omega_{\mathrm{PB}}$ is the primary beam field of view, X and Y are conversion factors from angle and frequency to comoving distance, and $<..>$ denote the averaging over baselines. The Fourier modes are in units of inverse comoving distance and are given by~\citep{Morales06,Trott12}:
\begin{align}
&k_{\perp} = \frac{2 \pi |\mathbfit{u}|}{D_M(z)},\\ 
&k_{\parallel} = \frac{2 \pi H_0 \nu_{21} E(z)}{c(1+z)^2} \tau,\\
&k = \sqrt{k_{\perp}^2 + k_{\parallel}^2},
\end{align}
where $D_M(z)$ is the transverse co-moving distance, $H_0$ is the Hubble constant, $\nu_ {21}$ is the frequency of the hyperfine transition, and $E(z)$ is the dimensionless Hubble parameter~\citep{Hogg10}. We also define the spherically-averaged power spectrum by averaging the power spectrum in spherical shells as:
\begin{equation}
\Delta^2({k}) = \frac{k^3}{2 \pi^2} P(k).
\end{equation}
The latter is well suited for characterising the 21-cm signal, which we expect to be isotropic at a given redshift. We limit our analysis to a bandwidth of 12 MHz to limit the effects of signal evolution, which would introduce anisotropy~\citep{Mertens20}.

To compute the power spectrum and its uncertainty for a specific component, say the 21-cm component, we start by taking $m$ samples from the posterior distribution of the hyperparameters, that we can take directly from the MCMC chain. We then generate ${\cal E}(\mathbf{f}_{\mathrm{21}})$ and $\mathrm{cov}(\mathbf{f}_{\mathrm{21}})$ using \autoref{eq:gpr_predictive_mean_eor} and \autoref{eq:gpr_predictive_cov_eor}, and produce the power spectrum of (${\cal E} + \delta_{21}$) with $\delta_{21}$ a sample drawn from the Gaussian distribution with covariance $\mathrm{cov}(\mathbf{f}_{\mathrm{21}})$. We then compute the power spectrum and associated 1-$\sigma$ uncertainty of the 21-cm component by taking the median and standard deviation of the $m$ power spectra thus generated.

\begin{figure}
    \includegraphics[width=\columnwidth]{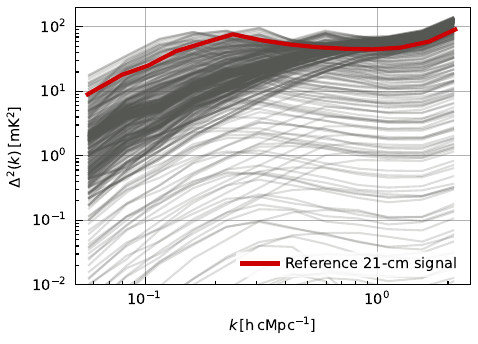}
    \caption{Spherically-averaged power spectra of 450 selected 21cmFAST simulations from the training set. The total training set is made of 5000 simulations. In red is shown the power spectra of the reference signal.}
    \label{fig:21-simulations}
\end{figure}

\section{Simulated data}
\label{sec:simulations}

\begin{figure*}
    \includegraphics[width=2\columnwidth]{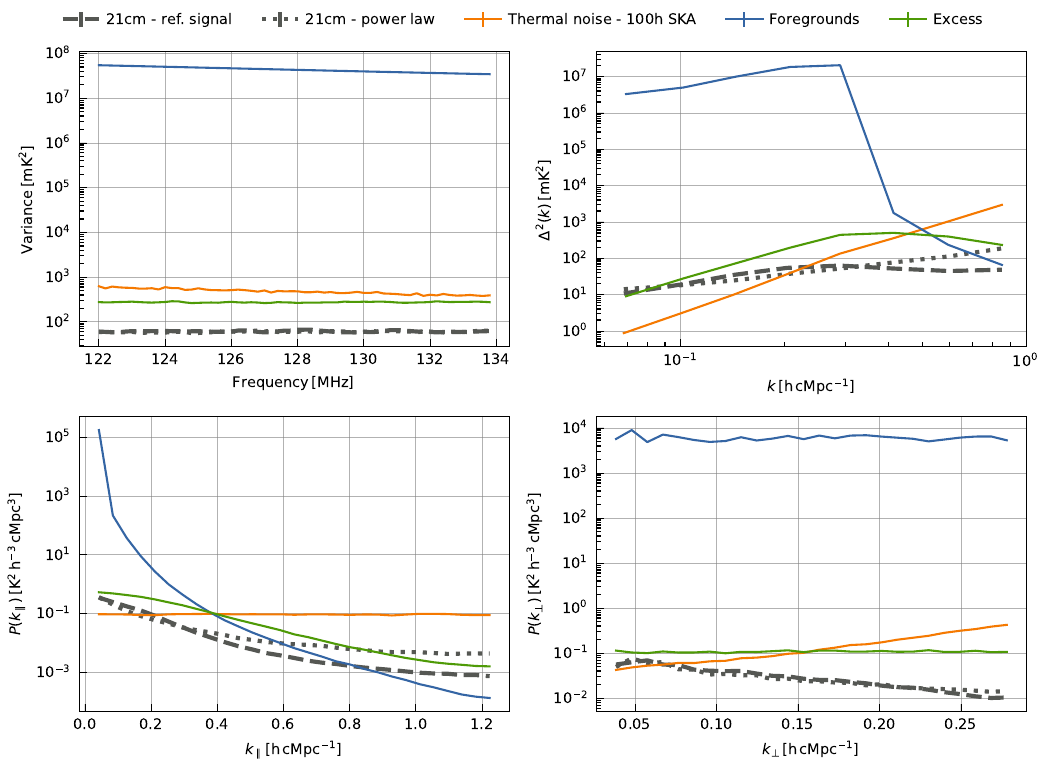}
    \caption{Power spectra of all simulated components used in our tests: foregrounds (blue), SKA thermal noise for 100h of observation (orange), excess component (green), reference 21-cm signal (dashed black line), synthetic 21-cm signal (dotted black line). We show their variance as function of frequency (top-left panel), the spherically-averaged power spectra (top-right panel), the cylindrically-averaged power spectra as function of $k_{\parallel}$ (bottom-left panel) and as function of $k_{\perp}$ (bottom-right panel).}
    \label{fig:ps_all_components}
\end{figure*}

This section provides a description of the simulated data sets used for testing the new ML-GPR method. We focus on SKA-low and simulate a gridded visibility cube using the planned SKA array configuration and the expected SKA thermal noise level. Noise simulation is performed using \texttt{ps\_eor}\footnote{\href{https://gitlab.com/flomertens/ps_eor}{https://gitlab.com/flomertens/ps\_eor}}. To produce the 21-cm signal component, we use 21cmFAST~\citep{Mesinger07,Mesinger11} simulated cubes for both the training set of the learnt kernel and the reference signal that we will attempt to recover. To model the foreground component, we use a similar strategy as in~\cite{Mertens18} and use Gaussian Processes (GP) with a frequency-frequency covariance model expected for a SKAO observation. As our primary focus is on validating the 21-cm learnt kernel, we use a simple foreground simulation generated with the same GP covariance function used for the signal separation to avoid bias from a mismatched foregrounds kernel : the same covariance model will be used as a foreground component in our GP model. However, we plan to try more realistic foreground simulations in future work. 

The simulation spans a frequency range of $122 - 134$ MHz with a spectral resolution of 0.2 MHz, corresponding to a redshift $z\sim10.1$. The maps cover a field of view of 4.5 degrees with a baseline ranging from $35\lambda$ to $300\lambda$, which gives us a maximum angular resolution of 10 arcmin. We chose these parameters to get an optimal sensitivity to the 21-cm signal.

\subsection{Foregrounds}
\label{sec:simulations_FGs}

Following \cite{Mertens18}, we model foregrounds using two components: the intrinsic astrophysical foregrounds and the instrumental effects, also known as the mode-mixing component.

At the radio frequencies at which 21-cm experiments operate, the astrophysical foregrounds emission foregrounds can be classified into two primary groups: (i) galactic foregrounds, which are primarily linked to the diffuse synchrotron and, to some degree, free-free emission from the Milky Way; and (ii) extragalactic foregrounds, which are linked to radio emission from star-forming galaxies and Active Galactic Nuclei, and to a lesser extent, radio halos and relics. The emission mechanisms are well known and are characterised by a power-law spectrum. Our approach in this work is not to model the spatial distribution of astrophysical foregrounds. Instead, we simulate this component using a Gaussian covariance with an 80 MHz frequency-frequency coherence scale. We do so because the ML-GPR method only accounts for frequency-frequency correlations, not spatial ones. Hence, our treatment is appropriate, given the limitations of our current implementation. The variance of this component is matched to the variance observed in LOFAR observations after Direction-Dependent calibration and sky model subtraction~\citep[see Figure 7 in][]{Mertens20}.

We use a Gaussian Process to simulate instrumental effects, which can arise from various sources, such as instrument chromaticity and calibration imperfections. Our approach is motivated by the analysis of LOFAR data, which suggests that medium-scale fluctuations can be accurately modelled using a Matérn covariance function,  as previously done in~\cite{Mertens18}. The mode-mixing typically appears in $k_{\perp}$,$k_{\parallel}$-space as a wedge-like structure~\citep{Datta10,Morales12}, with the level of spectral fluctuations increasing as the baseline length grows. To address this effect, we scale the frequency-frequency coherence scale with the angular scale, with values ranging from 4 MHz for a baseline length of $35\lambda$ to 2 MHz for a baseline length of $300\lambda$. Here we set $\eta = 5/2$ for the Matérn function.

In this work, we also investigate the resilience of our method to recover the 21-cm signal in the presence of additional, less spectrally-smooth components. Specifically, we consider an excess component observed in LOFAR observations~\citep{Mertens20} that cannot be attributed to astrophysical foregrounds, instrumental effects, or the 21-cm signal. To simulate this component, we use a Matérn kernel with $\eta=5/2$ and a coherence scale of 0.4 MHz, which captures its statistical characteristics. We aim to evaluate the robustness of our method in handling such contaminants.

\subsection{21-cm signal}

We simulate the 21-cm signal with 21cmFAST~\citep{Mesinger07,Mesinger11}, a semi-numerical code that characterises the relevant physical processes with approximate methods, resulting in faster and less expensive computations than accurate radiative transfer simulations. The semi-analytic codes generally agree well with hydrodynamical simulations for comoving scales $> 1 {~\rm  Mpc}$.

Several parameterisation options are available in 21cmFAST~\citep[e.g.,][]{park19}. In our study, we opt for the simpler three-parameter parameterisation, where $\zeta_0$ represents the ionising efficiency of high-$z$ galaxies, $T_\mathrm{vir}$ denotes the minimum virial temperature of star-forming haloes, in units of $M_{\odot}$, and $R_{\mathrm{mfp}}$ which represents the mean free path, in cMpc, of ionising photons within ionising regions. We produce a total of 5000 simulation co-eval cubes at $z\sim 10.1$, with $\zeta_0$ ranging between 10 and 250 with a step of 5, $\mathrm{log}(T_\mathrm{vir})$ ranging between 4 and 6 with a step of 0.1 and $R_{\mathrm{mfp}}$ ranging between 5 and 25 cMpc with a step of 2 cMpc. For the fiducial simulation, we use the parameters [$\zeta_0$, $\mathrm{log}(T_\mathrm{vir})$, $R_{\mathrm{mfp}}$] = [102, 4.65, 14], outside of the training set. In~\autoref{fig:21-simulations}, we plot in red the spherically-average power spectrum of the fiducial simulation against the spectra of some of the simulations from the training set (gray lines).

\subsection{Noise}

We test the signal separation algorithm in the context of a SKA-Low observation. With this aim, we simulate realistic SKA thermal noise using the \texttt{ps\_eor} code. At completion, the SKA-Low will be made of 512 stations, each of 256 elements, providing exquisite sensitivity at our scale of interests. Our simulation process involved creating the $uv$-coverage of the instruments by simulating the $uv$-tracks of each individual baseline and then gridding them to a $uv$-grid. The \textit{uv}-coverage grid is simulated assuming the current plan for antennae distribution of SKA-Low\footnote{The SKA-Low design is given at \url{https://www.skao.int/sites/default/files/documents/d18-SKA-TEL-SKO-0000422_02_SKA1_LowConfigurationCoordinates-1.pdf}.}. We simulate the $uv$-tracks corresponding to a 10-hour long observation, with an integration time $t_{\rm int} = 10$~s targeting the Murchison Widefield Array (MWA) EoR-0 deep field (Right Ascension = 0h00, Declination = $-27 \deg$), which is a potential future target for SKA-Low observations. The noise in each $uv$-cell, and for each frequency channel $\nu$ is then generated following a Gaussian distribution with standard deviation:
\begin{equation}\label{eq:syst_noise}
\sigma(u, v, \nu) = \frac{k_{\rm B}\, T_{\rm sys}(\nu)}{A_{\rm eff}} \sqrt{\frac{1}{2\delta_{\nu}N(u, v,\nu) t_{\rm int}}} .
\end{equation}

\subsection{Tests scenarios}

In \autoref{fig:ps_all_components}, we summarise all the components of our reference simulation and their differences in terms of variance (top-left panel), spherically-average power spectrum (top-right) and cylindrically-average power spectra averaged (bottom). We also include a different, {\it synthetic}, power-law like 21-cm signal that we will later employ in additional tests.

We assess the performance of the ML-GPR methods by testing them against various scenarios which are made of different components:

\begin{description}
    \setlength\itemsep{0.2em}

    \item[\bf Without Foregrounds:] This test will evaluate the raw performance of the algorithm with an ideal scenario including only the 21-cm signal and noise components.
    \item[\bf With Foregrounds: ] Evaluating the impact of foregrounds by including the 21-cm signal, noise, and foregrounds.
    \item[\bf Without 21-cm signal:] This will be a null test in which no 21-cm signal is present in the data, just noise and foregrounds.
    \item[\bf With an excess component:] In addition to the foregrounds, an extra component is added to the observation. In this case, we will test two possibilities: if the excess is accounted for or not in the GPR covariance model.
    \item[\bf Synthetic 21-cm signal:] Finally, the robustness of the methods is tested in recovering a class of models which are very different from what is used in the training set by employing a power-law like 21-cm signal.
\end{description}

For all these scenarios, we will compare the performance of the IAE and VAE kernels to the approach of \cite{Mertens18}, which uses an analytically defined covariance.

\begin{figure}
    \includegraphics[width=\columnwidth]{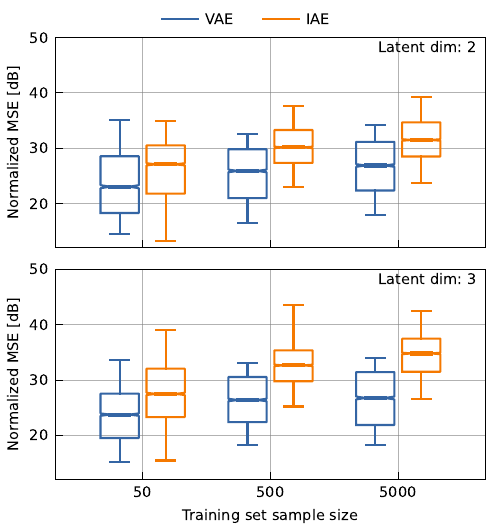}
    \caption{Normalized MSE distribution between the estimated and true power spectra from the test set for the VAE (orange) and IAE (blue) models, for $n=2$ (top) and $n=3$ (bottom).}
    \label{fig:vae_iae_MSE}
\end{figure}

\begin{figure}
    \includegraphics[width=\columnwidth]{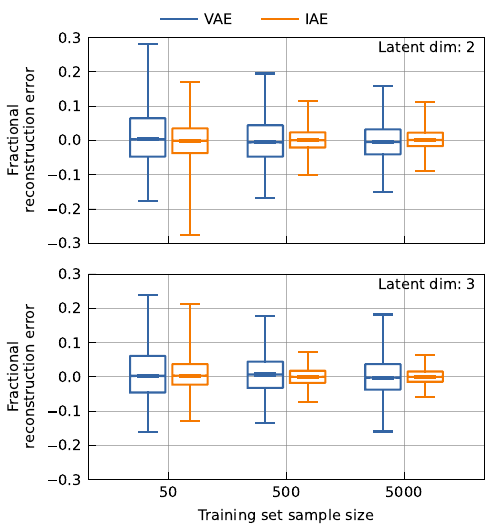}
    \caption{Normalized fractional ratio distribution between the estimated and true power spectra from the test set for the VAE (orange) and IAE (blue) models, for $n=2$ (top) and $n=3$ (bottom).}
    \label{fig:vae_iae_ratio}
\end{figure}

\section{Results}
\label{sec:results}

In this section, we first evaluate the reconstruction performance of the VAE and IAE and we later conduct various 21-cm signal recovery tests across different scenarios.

\subsection{Model evaluation}
\begin{figure*}
    \includegraphics[width=2\columnwidth]{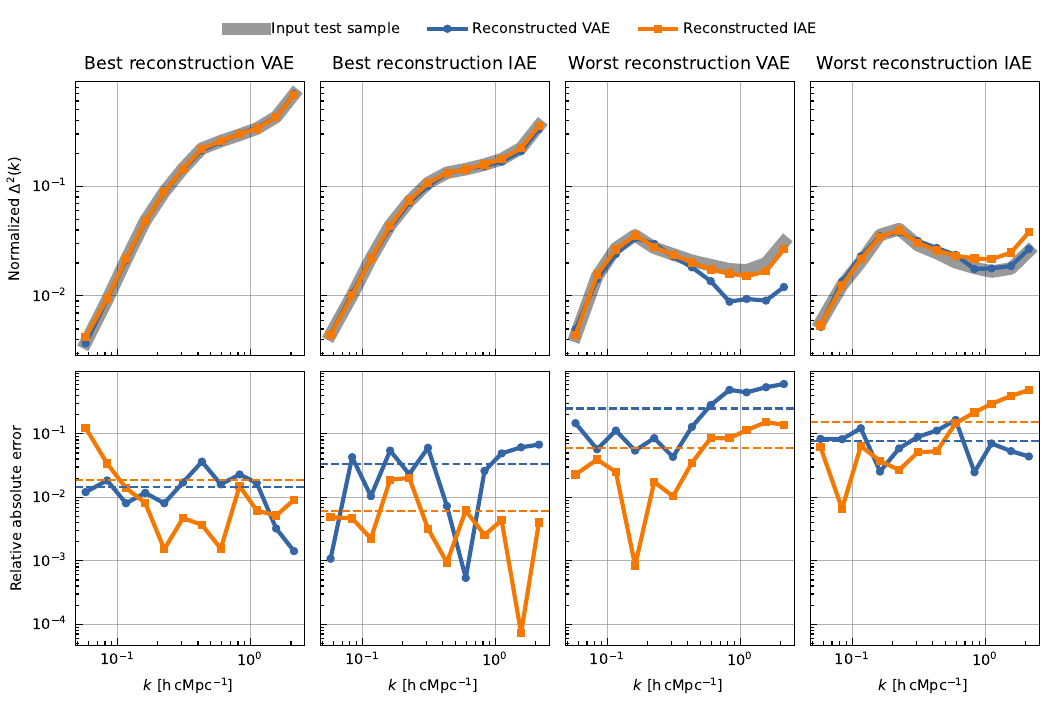}
    \caption{Reconstructed spherically-averaged power spectra (top) and residual (bottom) in the best VAE case (first panel), best IAE case (second panel), worst VAE case (third panel) and worst IAE case (fourth panel), estimated across the test set, showing both the VAE (blue) and IAE (orange) results for the four cases. For comparison, we show in gray the input signal. The averaged fractional absolute reconstruction error is also shown as a dashed line.}
    \label{fig:rec_best_averaged_worst}
\end{figure*}

The quality of the VAE- and IAE-based models are first evaluated independently of the regression task that we will eventually do. The goal is to assess the trained models' accuracy and ability to capture the structure of the $21$-cm power spectra. Whether it is IAE or VAE, both methods are used to build a low-dimensional and non-linear signal representation of the $21$-cm. To that purpose, these two approaches are evaluated with respect to two key parameters:
\begin{itemize}
\item The dimensionality $n$ of the latent space. $n$ sets the complexity of the trained models. An accurate low-dimensional model is important because a compressed representation allows for better discrimination between the $21$-cm signal and spurious contamination (e.g. foregrounds, noise, etc.).
\item The size of the training set. Evaluating the reliability of the trained model with respect to the size of the training set is crucial because the ability to train accurate models from scarce training samples will make possible the applications to more realistic EoR simulations, whose computational cost limits the number of samples for an eventual training stage.
\end{itemize}

\autoref{fig:vae_iae_MSE} displays the normalized mean squared error (NMSE) in decibels of the estimated $21$-cm signal power spectrum across the test set for latent space dimensionality $n=2$ and $n=3$. Comparing the top and bottom panels, and as expected, increasing the dimensionality improves the accuracy of trained models for the IAE and VAE, especially when the training set is large enough (e.g. the number of samples larger than $500$ in this experiment). For a small training set, the accuracy of the models is likely limited by the limited number of available training samples.

For $n=2$ and $n=3$, and as expected, both the IAE and VAE provide increasingly more accuracy as the number of training samples increases. The IAE provides more accurate models in all cases, with a significant NMSE gain close to $10$ dB for $n=3$ and a number of training samples of $5000$. The scatter of the NMSE across the test set is similar for both methods and decreases slightly when the number of training set samples increases.

The NMSE measures the discrepancy between the estimated and input power spectra that will be dominated by errors on its largest values located at large $k$ modes. The relative error ratio displayed in \autoref{fig:vae_iae_ratio} is more sensitive to the lowest values of the power spectra (i.e. at small $k$ modes). In this case, the VAE and IAE yield quite comparable average accuracy. Except for the case of $50$ training samples, the IAE yields slightly smaller dispersion across the test set. Considering the limited improvement gained by the higher dimensionality of the latent space and the increased complexity in the covariance model, we will focus on the $n=2$ case for the remainder of this paper. 

In \autoref{fig:rec_best_averaged_worst}, we present the reconstructed power spectra corresponding to the best and worst cases of relative absolute errors for both VAE and IAE. For these experiments, we use the model trained with 5000 samples and $n=2$. Overall, the IAE tends to produce more accurate low-dimensional models of the $21$-cm power spectra than the VAE. Even when the VAE exhibits the best reconstruction performance, the IAE performs similarly. Moreover, the relative absolute error of the worst IAE case is significantly lower than that of the worst VAE case. In general, The VAE and IAE generally perform better when dealing with "monotonous" power spectra, while their performance degrades  when confronted with power spectra that contain more complex features. The next sections will focus on extending these comparisons in the context of power spectra estimation from noisy and contaminated signals.

\subsection{21-cm signal power spectrum estimation}

We now evaluate the performance of the VAE and IAE models in retrieving the 21-cm EoR signal from a contaminated data cube and compare them with the standard GPR method (using a Matern covariance for the 21-cm component with $\eta = 3/2$. We use the VAE and IAE models trained with dimensionality $n = 2$ and a training set of 5000 samples. We will refer to the ML-GPR method employing the VAE model as VAE-GPR. Similarly, the ML-GPR method using the IAE model will be IAE-GPR.

\subsubsection{Scenario I: 21-cm signal and noise}

\begin{figure*}
    \includegraphics[width=2\columnwidth]{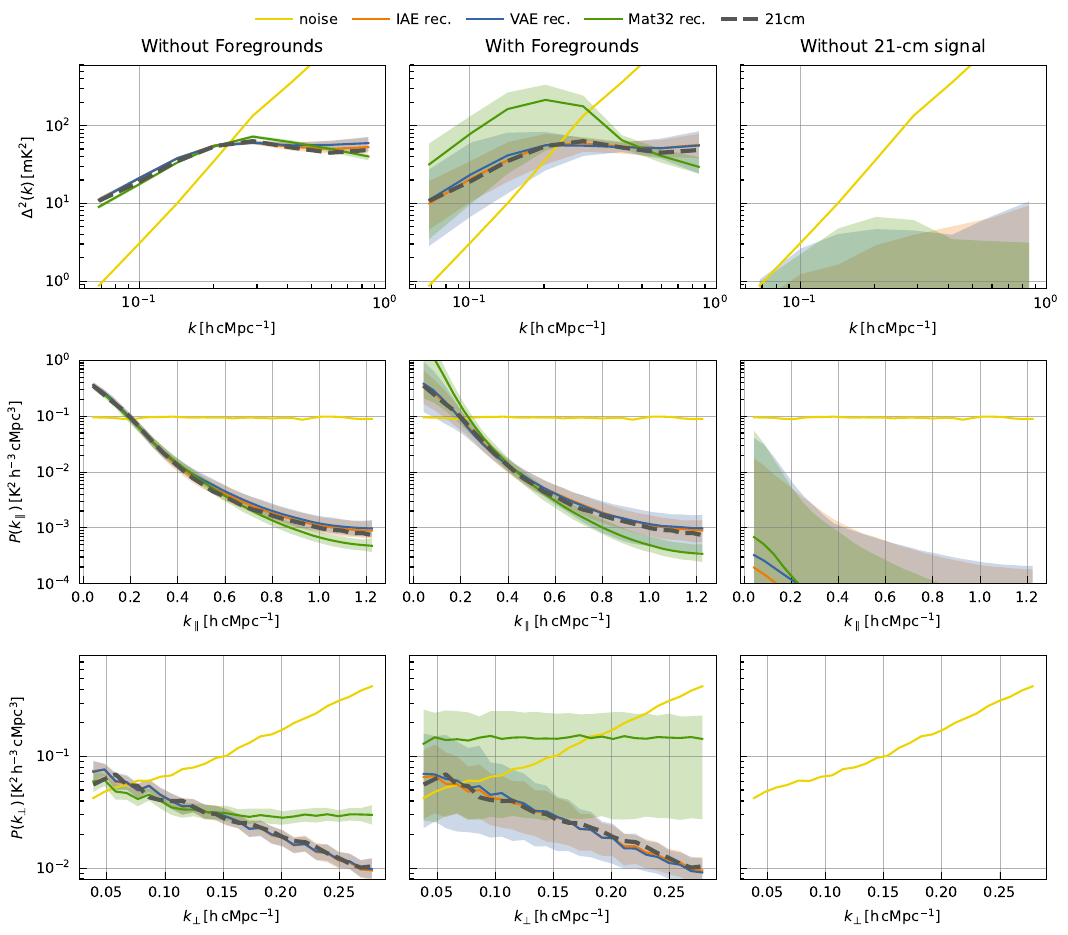}
    \caption{
    Recovered 21-cm signal power spectra using different methods: GPR (green), VAE-GPR (blue), and IAE-GPR (orange). The left panel shows the results when the data consist of only noise and the 21-cm signal. In the middle panel, a foregrounds component is added to the data, and in the right panel, the 21-cm component is absent. The power spectra of the noise component are shown in yellow, and the dashed black line represents the power spectra of the 21-cm component. The shaded area represents the 2-$\sigma$ uncertainty on the recovered 21-cm signal. The top panel displays the spherically-averaged power spectra, while the center and bottom panels shows the cylindrically-averaged power spectra as a function of $k_{\parallel}$ and  $k_{\perp}$.
    }
    \label{fig:iae_21cm_noise}
\end{figure*}

We begin by considering a simple scenario involving a data cube comprising only instrumental noise and the 21-cm signal. The results of the reconstruction are presented in the left panels of \autoref{fig:iae_21cm_noise}, where we compare the spherically-averaged power spectra (top panel) and the cylindrically-averaged power spectra along the line-of-sight (middle panel) and perpendicular to it (bottom panel). Even in this simple setup, both the VAE-GPR (in blue) and IAE-GPR (in orange) methods outperform the traditional GPR (in green), effectively recovering the input 21-cm signal. The GPR approach tends to underestimate the power at larger scales while overestimating it at smaller scales, particularly evident in the perpendicular direction. Interestingly, the scale (or $k_{\perp}$) dependence inherent in the new methods proves advantageous, contrasting with the lack of scale dependence in standard GPR, which result in overfitting at large and underfitting at small angular scales. Moreover, we report minimal uncertainties in the reconstruction, with slightly larger uncertainties at smaller scales (larger $k$) due to the lower signal-to-noise ratio in those regions.

\begin{figure*}
    \includegraphics[width=2\columnwidth]{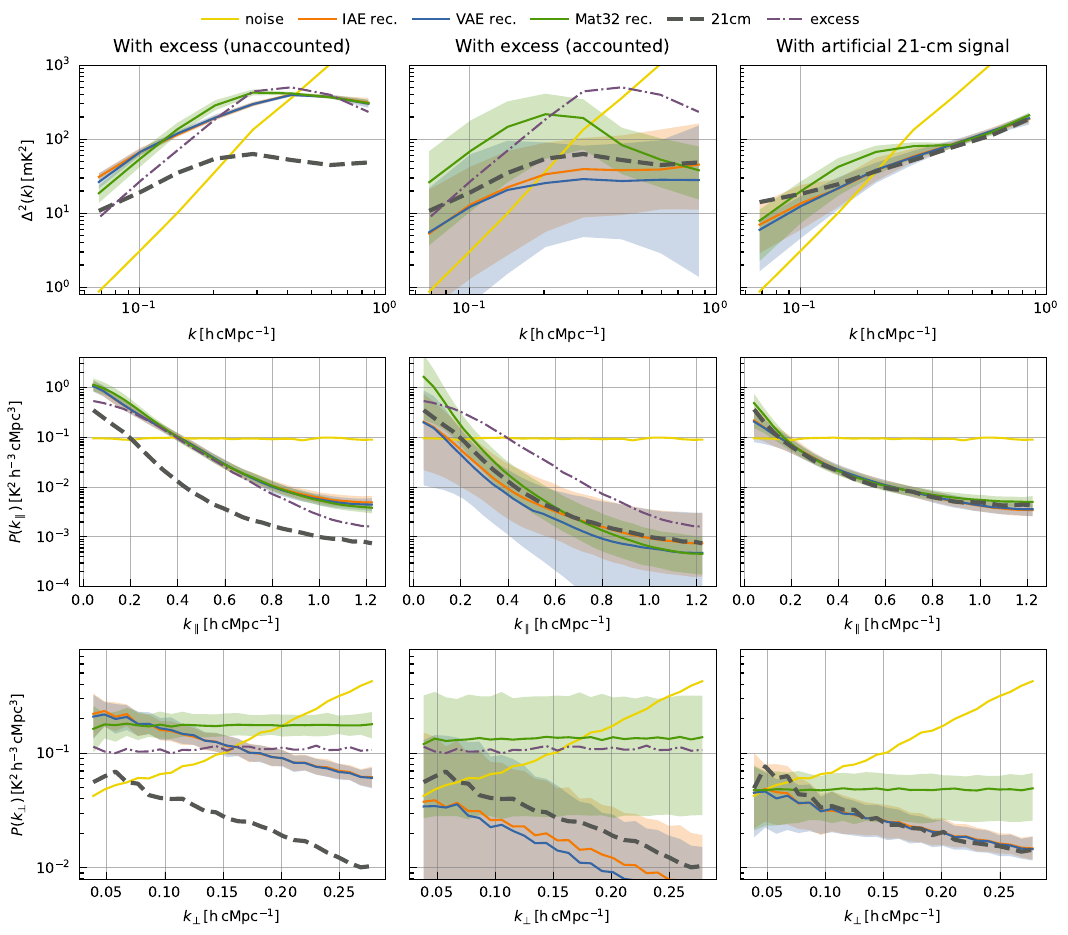}
    \caption{
    Recovered 21-cm signal power spectra using different methods: GPR (green), VAE-GPR (blue), and IAE-GPR (orange). In the left panel, an excess component is introduced into the data without being accounted for in the GPR model. The middle panel demonstrates the results when the excess component is properly accounted for. In the right panel, the reference 21-cm signal is replaced with a synthetic 21-cm signal modeled by a power-law power spectra. The power spectra of the noise component are shown in yellow, and the dashed black line represents the power spectra of the 21-cm component. The shaded area represents the 2-$\sigma$ uncertainty on the recovered 21-cm signal. The top panel displays the spherically-averaged power spectra, while the center and bottom panels shows the cylindrically-averaged power spectra as a function of $k_{\parallel}$ and  $k_{\perp}$.
    }
    \label{fig:iae_21cm_noise_frg_cmpt2}
\end{figure*}

\subsubsection{Scenario II: 21-cm signal, noise and foregrounds}



In the middle column of \autoref{fig:iae_21cm_noise}, we consider the scenario where the data cube comprises the 21-cm signal, noise, and foregrounds, representing the ideal case without additional systematics in the data. Despite the presence of larger uncertainties, both the VAE-GPR and IAE-GPR methods successfully recover the 21-cm signal. In contrast, the standard GPR approach exhibits a worsened deviation from the ground truth. Notably, the limitations of the standard GPR arise from its inability to accurately capture the shape of the input power spectra, primarily due to the lack of baseline dependence of the covariance prior model. The latter is particularly evident when considering the power spectrum as a function of $k_\perp$ (bottom panel), which presents a flat behaviour. In contrast, the ML-GPR methods perfectly recover the shape of the input power spectra. The uncertainty is highest in regions dominated by foregrounds ($k < 0.3\,{\rm h\,cMcp^{-1}}$) or noise ($k > 0.5\,{\rm h\,cMcp^{-1}}$). Comparing the performance of the IAE and VAE, both methods exhibit equivalent recovery, although the IAE demonstrates a slightly smaller bias.



\subsubsection{Scenario III: noise only}

In this particular scenario, we evaluate the performance of the algorithm when the 21-cm signal is significantly below the sensitivity threshold of the instrument, resulting in a data cube consisting only of foregrounds and noise components. When looking at the result of the MCMC, in all three cases (VAE-GPR, IAE-GPR and standard GPR) the hyperparameters that control the shape of the power-spectra (the two latent space parameters for the VAE and IAE, and the Matern coherence-scale for the standard GPR) are unconstrained: there posterior is flat. The variance of this component is also unconstrained with a large right-tail which is set by the sensitivity (all variance hyperparameters are inferred in log-space). This is the reason why, also the recovered power-spectra in the left panel of \autoref{fig:iae_21cm_noise} is below the noise power-spectra (yellow line), the corresponding 2-sigma uncertainty is large and reach region which can be just below the noise. This is the case for all three methods.



\subsubsection{Scenario IV: 21-cm signal, noise, foregrounds and extra component}

To assess the robustness of our algorithm against more realistic systematic effects, we introduce an additional foreground component into the data cube, as described in Section 4.2. This component is designed to mimic the statistical properties of the "excess" observed in current LOFAR-EoR analysis~\citep{Mertens20}. Notably, this excess component is currently not subtracted from LOFAR data, making it a limiting factor in current analysis. While improved calibration and processing techniques may potentially mitigate or eliminate this excess component, the ability to robustly separate it from the 21-cm signal is crucial for maximizing the experiment's sensitivity.

In the initial analysis, we examine the recovered 21-cm signal without including any additional component in our prior covariance model, considering only the foregrounds, noise, and 21-cm component. The results presented in the first column of \autoref{fig:iae_21cm_noise_frg_cmpt2} indicate poor performance for all three methods. We find that the recovered signal comprises both the excess component and the 21-cm signal. The excess present in the data is being absorbed by additional variance in the 21-cm component. Since this excess component was not accounted for in the covariance prior model, the algorithm effectively attributes the unexplained variance to the 21-cm signal. However, we also observe a discrepancy in the shape of the recovered signal power spectra compared to the shape of the excess power spectra. This disparity arises from the fact that the covariance prior model for the 21-cm signal does not include this specific class of shape.

To address this limitation, we introduce a second step in our analysis where the same Matern covariance used to simulate the excess component is added to the GPR covariance prior model. By doing so, we account for the statistical properties of the excess component. This assumes that we have a thorough understanding of the systematic and can perfectly account for it in our analysis. The results (middle panel of \autoref{fig:iae_21cm_noise_frg_cmpt2}) demonstrate that while the uncertainty in the power spectra significantly increases, the input signal lies well within the 2-sigma uncertainty range of the recovered signal. Both IAE-GPR and VAE-GPR show successful recovery, while standard GPR performs significantly worse. We observe a greater mixing between the excess and 21-cm components in standard GPR, resulting in a stronger bias. This discrepancy is particularly pronounced when examining the power spectra in the $k_\perp$ direction, highlighting the importance of properly accounting for all components, including systematics, in the covariance prior model.

\subsubsection{Scenario V: synthetic 21-cm signal, noise and foregrounds}

In the last test scenario, we aim to recover a completely synthetic and non-physical model, which represents a case where the true power-spectra significantly deviate from current model predictions, and is thus not present in the training set. To simulate this scenario, we adopt a power-law model with the equation $\Delta^2(k) = \sigma_{21}^2 k^{2}$, where $\sigma_{21}^2$ is chosen to match the variance of the fiducial 21-cm signal. The results of this test, presented in the last column of \autoref{fig:iae_21cm_noise_frg_cmpt2}, reveal that while none of the methods are able to perfectly recover the 21-cm signal, the IAE-GPR and VAE-GPR methods exhibit better recovery of the shape compared to the standard GPR method. This improvement is particularly evident for $k$ modes greater than $0.1 , \mathrm{h , cMpc^{-1}}$. Upon examining the power spectra as a function of $k_{\parallel}$, we observe that the discrepancy between the input and recovered signal is more pronounced for modes where the foregrounds dominate, specifically the low $k_{\parallel}$ modes.

This test underscores the importance of having a sufficiently wide prior for the 21-cm signal in order to account for unexpected 21-cm signal. It also highlights the need for a feedback loop with theoretical models, where simulations can be updated to explain observations, signal separation can be re-run with the updated models, and the resulting signal can be refined accordingly.

\subsubsection{Effect of varying the noise level}

\begin{figure}
    \includegraphics[width=\columnwidth]{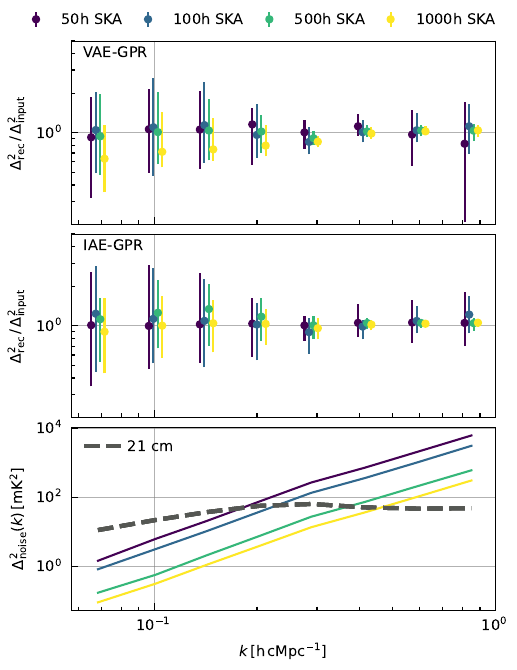}
    \caption{
    Ratios of recovered over input 21-cm power spectra using a VAE (top panel) and IAE (middle panel) trained with 5000 samples and for different level of noise in the data: between 50h and 1000h of equivalent SKA noise. Each case is depicted with a small offset around the measured $k$-bin for clarity. The 2-sigma uncertainty is also indicated for each case. In the bottom panel, the power spectra of the noise for the different cases are shown, along with the power spectra of the 21-cm signal, illustrating the signal-to-noise ratio (SNR) due to the noise. Foregrounds dominate the power spectra for $k < 0.3\,{\rm h\,cMcp^{-1}}$.
    }
    \label{fig:varying_intensity}
\end{figure}

In this section, we explore the influence of varying thermal noise levels on the recovery of the 21-cm power spectra. To investigate this, we conduct simulations with different integration times, ranging from 50 to 1000 hours. The results of this analysis are presented in Figure \ref{fig:varying_intensity}, illustrating the ratio between the recovered and input 21-cm power spectra using the VAE and IAE-trained kernel.

Our findings indicate that reducing the thermal noise level has minimal impact on the uncertainty of the recovered signal for the $k$-modes dominated by foregrounds ($k < 0.3,{\rm h,cMcp^{-1}}$). This suggests that the primary source of uncertainty in the foreground-dominated region stems from the foreground itself rather than the noise. However, for $k$-modes where noise dominates ($k > 0.4,{\rm h,cMcp^{-1}}$), increasing the integration time substantially decreases the uncertainty. These results underscore the crucial role of foregrounds in contributing to the uncertainty in the recovery of the 21-cm power spectra.

\subsubsection{Varying the size of the training set}

In this section, we examine the influence of the training set size on the recovery of the 21-cm signal using our trained kernel. While gathering a large training set is feasible for semi-numerical simulation codes like 21cmFast, it may become computationally prohibitive for full 3D radiative transfer simulation codes \citep[e.g.,][]{Ciardi01,Semelin17}. However, the generative nature of our methods enables us to achieve satisfactory recovery even with a   relatively small dataset.

\begin{figure}
    \includegraphics[width=\columnwidth]{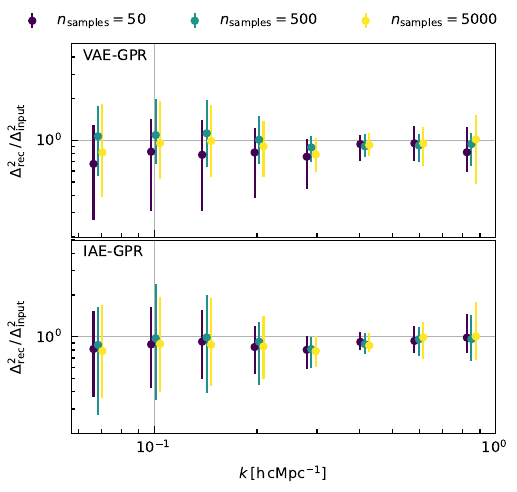}
    \caption{
    Ratios of recovered over input 21-cm power spectra using a VAE (top panel) and IAE (bottom panel) with different numbers of samples in the training set, ranging from 50 (blue) to 5000 samples (yellow). Each case is depicted with a small offset around the measured $k$-bin for clarity. The 2-sigma uncertainty is also indicated for each case.}
    \label{fig:varying_nsamples}
\end{figure}

We assess the behaviour of the recovery using different training set sizes, and the results are presented in Figure \ref{fig:varying_nsamples}. Remarkably, even with only 100 samples, we can still recover the 21-cm signal, albeit with higher bias and uncertainty. This effect is predominantly observed in the region dominated by foregrounds ($k < 0.3,{\rm h,cMcp^{-1}}$), while the noise-dominated region remains relatively unaffected. Notably, the case with 500 samples performs nearly as well as that with 5000 samples.

These findings demonstrate the capability of our methods to achieve reliable recovery of the 21-cm signal, even with limited training data. While a larger training set generally leads to improved performance, our approach proves robust in cases where computational constraints limit the availability of extensive training samples.

\section{Discussion \& Conclusions}







Detecting the redshifted 21-cm radiation of neutral hydrogen is the ultimate probe of the early stages of the observable Universe (Dark Ages and Epoch of Reionization). Even though a tremendous observational effort is ongoing --culminating with the SKA Observatory being online by the end of this decade-- these observations are hard to perform as the signal is dwarfed by astrophysical contaminants shining at the same wavelengths, which are coupled to intricate systematics. Among the strategies developed to address these challenges, we find GPR, introduced in the 21-cm context by \cite{Mertens18} and successfully applied to LOFAR, AARTFAAC and HERA data~\citep{Mertens20,Gehlot19,Gehlot20,Ghosh20}. 

The scope of this work has been improving GPR, making it more adaptable to the component separation task to retrieve the 21-cm signal. In particular, standard GPR assumes that the frequency-frequency covariance of the different components can be expressed analytically. We substituted that framework with covariance models learnt from 21-cm signal simulations. To do so, we relied on two different autoencoders: variational and interpolatory autoencoders. We tested the three methods (GPR, VAE-GPR and IAE-GPR) in various scenarios to assess their ability to pinpoint the accurate power spectra of the 21-cm signal.

Firstly, we demonstrated that using $n=2$ latent variables is sufficient to characterise the spherically-averaged power spectrum of the 21-cm signal using both tha VAE and IAE schemes based on our study's specific 21cmFast simulation set. Increasing the number of samples in the training sets enhances the reconstruction capability of the autoencoders, resulting in a representation that better reflects the true signal. However, it is important to note that this improvement in training does not necessarily translate to better inference when recovering the 21-cm signal from a simulated observation. Generally, the AE reconstruction is more successful for power spectra with flatter profiles compared to those with more complex features. In the best cases, both VAE and IAE perform similarly. Still, VAE is more prone to mismodel power spectra with intricate features.

We then demonstrated the advantages of employing ML-GPR for component separation tasks by evaluating various scenarios. The inherent $k_\perp$ dependence in the learned kernel proved significant, distinguishing it from the standard GPR approach, which lacks this information. In general, both VAE and IAE-based kernels exhibited improved recovery of the 21-cm signal power spectra, characterized by reduced bias and lower uncertainty than standard GPR. This can be attributed to the learnt kernel being constrained to a set of covariances that align with our expectations for the 21-cm signal. 

Promisingly, when an additional contaminant, simulating a systematic in the data, was introduced to the data cube, both VAE-GPR and IAE-GPR demonstrated their effectiveness. When the systematic was properly accounted for in the prior covariance model, these methods successfully recovered the power spectra of the 21-cm signal, although with higher uncertainty. Moreover, even when the systematic was not accounted for, VAE-GPR and IAE-GPR still provided a correct upper limit on the power spectra of the 21-cm signal, by not over-fitting the unknown component. This highlights the robustness of VAE-GPR and IAE-GPR in the presence of systematics, showcasing their capability to deliver accurate signal estimates in challenging scenarios.

Significantly, the two ML-GPR outperform GPR in the scenario where we use a 21-cm signal not represented in the training set (nor in the GPR analytical kernel framework). The latter test is essential before applying ML-GPR in a real observational context, as it assesses the importance of the unknown representativeness of the training set.

In general and expectedly, uncertainties are higher in the $k$-regions where foregrounds or noise are dominant. When increasing the total observing time, we observed a decrease in uncertainty, as anticipated, in the noise-dominated region. However, the decrease in uncertainty was only marginal in the foreground-dominated region. We also found that for this specific 21cmFast simulation set, the number of samples in the training set does not play a significant role in the component separation task.

As for the differences between the two new algorithms, IAE-GPR tends to have a slightly higher fidelity recovery of the signal power spectra than VAE-GPR. We remind that VAE looks after a global description of the manifold where the 21-cm power spectra live, tending to generalise. However, IAE works with true anchor points on the manifold from which the algorithm can interpolate the output power spectra, making it more flexible in the reconstruction.

For real applications in the short term, we expect the different performances of the two ML-GPR to be  insignificant as we deal with highly contaminated and noise-dominated data. Nonetheless, the IAE architecture is robust also in the case of few available training samples, which, in the short term, can allow us to train it with 21-cm simulations that are more realistic than what we have exploited here, but also more computationally expensive and thus more scarce. 

The 21-cm from the Epoch of Reionization and Cosmic Dawn still needs to be better understood; many simulation codes are integrating more complex physics. We should combine multiple simulations code and train on a combined set to widen the prior of the 21-cm component and cover all possibilities.

As another future outlook, we plan to further test these methods' robustness against more realistic contaminants. The test with the excess component (Scenario IV, left column of \autoref{fig:iae_21cm_noise_frg_cmpt2}) shows that in the presence of an unaccounted component, its contribution to the variance in the data will be absorbed by other components --most likely by the 21-cm part which is flexible enough. As such, constructing the correct covariance prior model that can account for all data systematics is of utmost importance, informed in this stage by solid statistical techniques. 

Summarising, the critical outcomes of our work are that GPR with a trained kernel can recover a 21-cm signal that was not described in the training set and, to some extent, can do so also when an unknown component --which we will not be able to simulate-- contaminates the data cube. Our results show the strength of these generative approaches and represent a further step forward in optimising the analysis techniques that will make the detection at $z>6$ of the 21-cm signal possible.

\section*{Acknowledgements}

FGM acknowledges support of the PSL Fellowship. IPC is supported by the European Union within the Next Generation EU programme [PNRR-4-2-1.2 project No. SOE\_0000136].

\section*{Data Availability}

The code to reproduce our results, as well as the 21cmFast training set and a step by step procedure, will be openly available upon publication.


 



\bibliographystyle{mnras}
\typeout{} 
\bibliography{references} 




\appendix

\section{Implementation details about the VAE}
\label{app:details}
Table~\ref{table:params} displays the details of the architecture of both the IAE and VAE used in this article.

\begin{table}
\begin{tabular}{|lcc}
\hline
 & VAE & IAE \\
 \hline
  \hline
Nb. of layers & 3 & 3 \\
 \hline
Encoder & [20,20,20]  & [12,12,12]  \\
 \hline
Decoder & [20,20,20] & [12,12,12] \\
 \hline
batch size & 128 & 64\\
 \hline
epochs & 2000 & 5000 \\
 \hline
beta & 1e-4 & 1e-3 \\
 \hline
latent dim &  2 & 3 \\
 \hline
learning rate & 0.02 & 0.01 \\
 \hline
learning rate decay & $1e-5$ & none \\
 \hline
\end{tabular}
\label{table:params} \caption{Details of the architecture of both the IAE and VAE used in this work.}
\end{table}

\bsp	
\label{lastpage}
\end{document}